\documentclass[runningheads]{llncs}
\usepackage[T1]{fontenc}
\usepackage{graphicx}
\usepackage{amsmath}
\usepackage{multirow}
\usepackage{amssymb}
\usepackage{bm}
\usepackage{ulem}
\usepackage[colorlinks]{hyperref}

\usepackage{cleveref}
\usepackage{booktabs}
\usepackage{marvosym}

\begin{document}
\title{3DGR-CAR: Coronary artery reconstruction from ultra-sparse 2D X-ray views with a 3D Gaussians representation}
\titlerunning{3DGR-CAR}

%

\authorrunning{X. Fu et al.}

\author{Xueming Fu\inst{1,2} 
\and Yingtai Li\inst{1,2} 
\and Fenghe Tang\inst{1,2} 
\and Jun Li\inst{1,2} 
\and Mingyue Zhao\inst{1,2} 
\and Gao-Jun Teng\inst{5} 
\and S. Kevin Zhou\inst{1,2,3,4} 
$^{\href{mailto:s.kevin.zhou@gmail.com}{\textrm{\Letter}}}$}

\institute{School of Biomedical Engineering, Division of Life Sciences and Medicine, University of Science and Technology of China (USTC), Hefei Anhui, 230026, China \and
Center for Medical Imaging, Robotics, Analytic Computing \& Learning (MIRACLE), Suzhou Institute for Advance Research, USTC, Suzhou Jiangsu, 215123, China \and
Key Laboratory of Precision and Intelligent Chemistry, USTC, Hefei Anhui, 230026, China \and
Key Laboratory of Intelligent Information Processing of Chinese Academy of Sciences (CAS), Institute of Computing Technology, CAS \and
Center of Interventional Radiology \& Vascular Surgery, Department of Radiology, Medical School, Zhongda Hospital,  Southeast University, Nanjing 210009, China 
}

\maketitle             
\begin{abstract}

Reconstructing 3D coronary arteries is important for coronary artery disease diagnosis, treatment planning and operation navigation. 
Traditional reconstruction techniques often require many projections, while reconstruction from sparse-view X-ray projections is a potential way of reducing radiation dose.
However, the extreme sparsity of coronary arteries in a 3D volume and ultra-limited number of projections pose significant challenges for efficient and accurate 3D reconstruction. To this end, we propose 3DGR-CAR, a 3D Gaussian Representation for Coronary Artery Reconstruction from ultra-sparse X-ray projections. We leverage 3D Gaussian representation to avoid the inefficiency caused by the extreme sparsity of coronary artery data and propose a Gaussian center predictor to overcome the noisy Gaussian initialization from ultra-sparse view projections. 
The proposed scheme enables fast and accurate 3D coronary artery reconstruction with only 2 views. Experimental results on two datasets indicate that the proposed approach significantly outperforms other methods in terms of voxel accuracy and visual quality of coronary arteries. The code will be available in https://github.com/windrise/3DGR-CAR.

\end{abstract}

\section{Introduction}
Cardiovascular disease, particularly coronary artery disease (CAD), is becoming increasingly prevalent worldwide \cite{CAD}. Accurate 3D reconstruction of coronary arteries greatly assists physicians in the diagnosis and treatment planning of CAD \cite{Invasive-CA,assist_doctor}, enabling them to make informed decisions and provide targeted interventions. 
However, many projections are required during angiography to obtain accurate spatial structures of the vessels.
%

\begin{figure}[t]
    \centering
    \includegraphics[width=.9\textwidth]{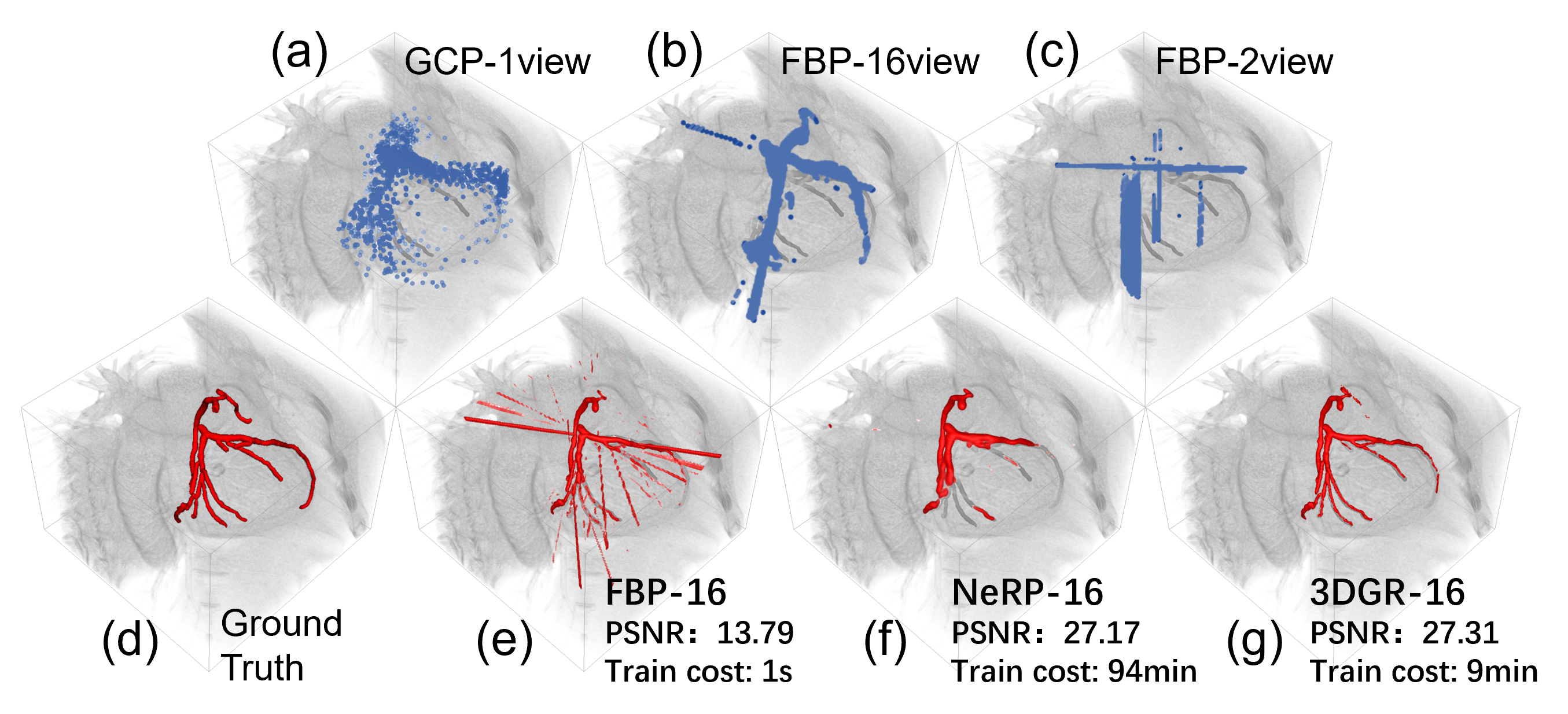}
    \caption{Coronary arteries occupy only approximately 0.1\% of a typical cardiac volume. (a-c): Gaussian centroid initial positions derived from our Gaussian center predictor and FBP results. (d): Ground truth. (e-g) Coronary reconstruction results from FBP \cite{fbp1976reconstruction}, NeRP \cite{shen2022nerp} and 3DGR.} \vspace{-10pt}
    \label{fig:intro}
\end{figure}

Classic scanning and reconstruction approaches (such as Feldkamp-Davis-Kress \cite{fdk1984practical} and Filtered Back-Projection (FBP) \cite{fbp1976reconstruction}) that depend on dense view data require patients to be continuously exposed to ionizing radiation \cite{Grass20003D}. In contrast, sparse-view reconstruction techniques, which utilize a limited number of projections to reconstruct 3D structures, can potentially reduce radiation dose and minimize the risk to patients while still providing valuable diagnostic information \cite{sidky2006accurate}. Given the latest technical advances and the fact that coronary arteries occupy only approximately 0.1\% of an entire volume for a typical cardiac scan (Fig.~\ref{fig:intro}), a question naturally arieses: {\bf is it possible to utilize a really sparse number of 2D X-ray views to reconstruct coronary arteries in 3D? }


Due to its independence from training data, the neural radiation field (NeRF) \cite{mildenhall2021nerf} approach holds a great potential for medical image reconstruction \cite{shen2022nerp,maas2023nerf-cA,zha2022naf,fang2022snaf}. 
Shen {\it et al.} propose an implicit neural representation (INR) with prior embedding (NeRP) to reconstruct images from sparsely sampled views \cite{shen2022nerp}. 
However, INR-based approaches struggle with coronary artery reconstruction (CAR) due to the extreme sparsity of coronary artery image, resulting in a slow speed and limited performance according to our empirical observation, which hinders their practical application in clinical settings.
Recently, the 3D Gaussian Splatting \cite{kerbl20233dGS} has emerged as a notable strategy for reconstructing 3D scenes from images, with superior quality and faster convergence compared to NeRF methods. 
We identify 3D Gaussian representation (3DGR) as particularly well-suited for reconstructing extremely sparse objects such as coronary arteries, as they can be initialized from a sparse set of point clouds outlining the object, thereby avoiding unnecessary computation in empty spaces. 
A few pioneering attempts have been made to introduce 3DGR to the medical imaging field \cite{li2023sparse-gs,zhu2024deformable-gs,liu2024endogaussian}. For example, 
Li {\it et al.} utilize FBP-reconstructed images for initializing Gaussian parameters, achieving superior performance in sparse-view CT reconstruction compared to neural field methods \cite{li2023sparse-gs}. However, when directly applying these strategies for coronary artery reconstruction, we empirically obseve that the performance falls short of expectations. 


We recognize that the performance gap lies in the deterioration of initialized Gaussians centers as a result of increasing sparsity of the object and reduced number of projections. Accurate initialization plays a crucial role in determining the final reconstruction quality. However, the initialization of Gaussian model parameters heavily relies on point cloud data obtained through Structure-from-Motion (SfM) techniques \cite{kerbl20233dGS} or voxel data from FBP \cite{li2023sparse-gs}. 
Thus, the noise level in the initial point cloud generated by SfM or FBP increases dramatically with a decreasing number of projections, leading to highly inaccurate Gaussian parameter initialization, which significantly degrades the quality of the final reconstruction, as shown in Fig. \ref{fig:intro}. 

%
In this paper, we propose {\bf 3DGR-CAR}, a \textbf{3D} \textbf{G}aussian \textbf{R}epresentation scheme for \textbf{C}oronary \textbf{A}rtery \textbf{R}econstruction from ultra-sparse 2D X-ray views. 
Our main contributions are as follows:
(1) We introduce 3D Gaussian representation for coronary artery reconstruction. 3D Gaussians are initialized from a set of sparse point clouds, which avoids computation in empty spaces and improves efficiency, making it well-suited for the sparse nature of coronary artery images. This is \textbf{the first attempt} that successfully applies 3D Gaussian to coronary artery reconstruction.
(2) We employ a U-Net to provide initialization of Gaussian centers. Point clouds estimated from ultra-sparse view are often extremely noisy and can harm the reconstruction performance of 3D Gaussians. By training the U-Net, we can leverage the knowledge stored in its parameter to place the initialized Gaussian centers in better locations, enhancing the reconstruction accuracy. 
(3) We conduct comprehensive evaluations on the ImageCAS and ASOCA datasets, demonstrating that our proposed 3DGR-CAR substantially surpasses current INR approaches and vanilla 3DGR method in terms of reconstruction quality with a considerably reduced time cost (Fig.~\ref{fig:intro}).

\section{Methods}
\begin{figure}[t]
    \centering
    \includegraphics[width=\textwidth]{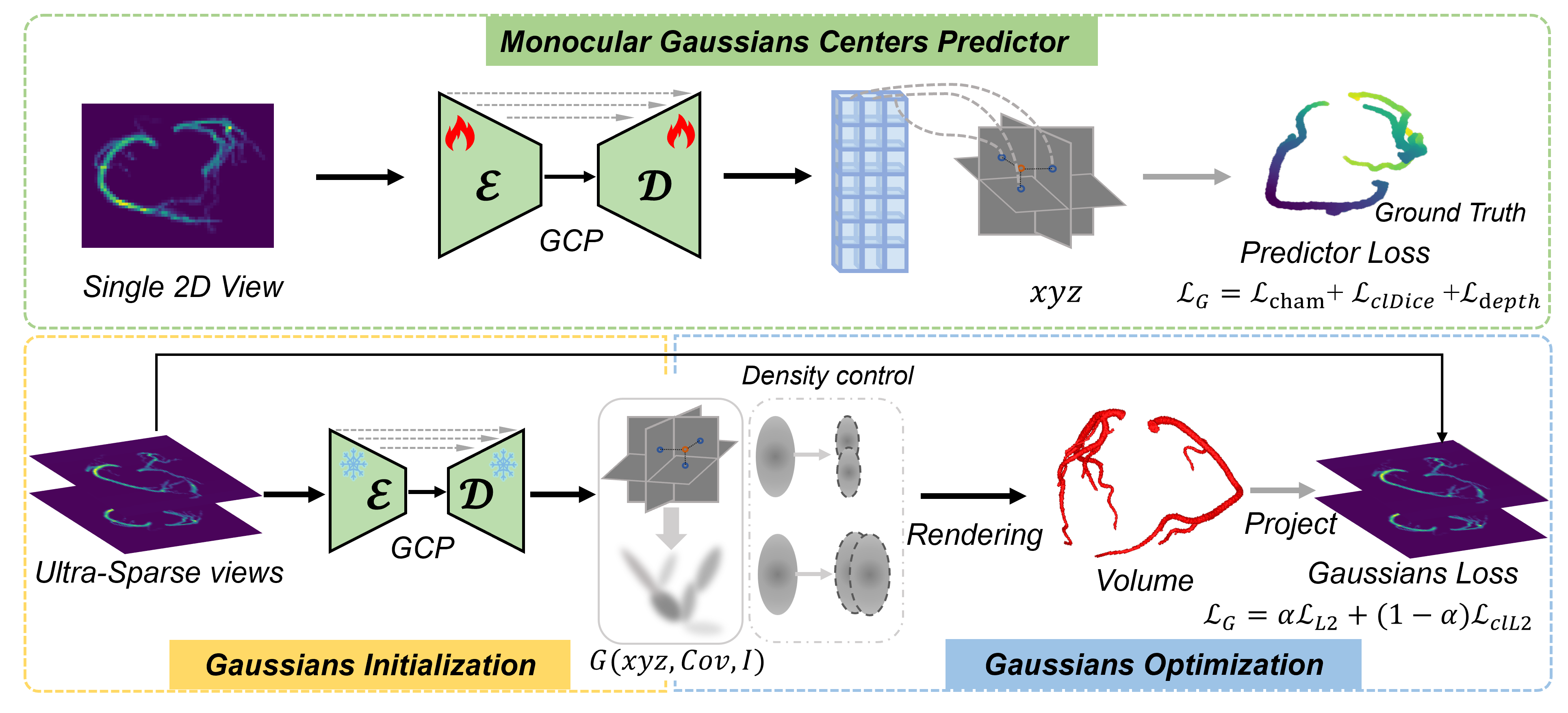}
    \caption{Overview of 3DGR-CAR. The green box illustrates the pipeline for training GCP; the yellow box denotes the Gaussian position are initialized by GCP; the blue box represents Gaussian parameters optimization with sparse-view projections.} \vspace{-10pt}
    \label{fig:framework}
\end{figure}

As in Fig. \ref{fig:framework}, our proposed 3DGR-CAR consists of two stages: the Gaussian Center Predictor (GCP) training stage and the 3DGR reconstruction stage. 
In the first stage, a U-Net network \cite{ronneberger2015u} is trained to estimate voxel depth from a single simulated artery X-ray projection.
In the second stage, a monocular image is processed through the U-Net to obtain the positional parameters \(M = (d, \Delta_{x}, \Delta_{y}, \Delta_{z})\). Subsequently, these predicted positional parameters are utilized for the initialization of 3D Gaussian centers. Afterwards, the parameters of the Gaussian model are optimized based on sparse vascular projection X-ray images, thereby reconstructing 3D coronary arteries from sparse views. In the following subsections, we first provide a brief overview of the 3DGR for CAR in Sec.~\ref{sec:overview}, then detail training of the Gaussian center predictor (in Sec.~\ref{sec:Predicting}). 

\subsection{3D Gaussians Representation Reconstruction}
\label{sec:overview}
3D Gaussian is a flexible and expressive scene representation \cite{kerbl20233dGS}. Gaussians are initialized from a set of sparse point clouds. Each 3D Gaussian is characterized by a set of parameters that define its position, shape, and other task-specific parameters. In this work, we align with \cite{li2023sparse-gs}, parameterize each Gaussian $G_i$ by $\theta_i = \{\mu_i, \Sigma_i, I_i\}$, where $\mu_i$ pinpoints the coordinate of the center position for $G_i$, $\Sigma_i$ is the covariance matrix that encapsulates information about the spread and orientation of the Gaussian, and $I_i$ is the intensity of $G_i$. Each 3D Gaussian contributes to the overall representation through a function that gauges its impact on any given point in space, with the influence of each Gaussian described by 
\begin{equation}
    G_i(X|\theta_i) = I_i \cdot e^{-\frac{1}{2}(X-\mu_i)^T \Sigma_i^{-1}(X-\mu_i)}, 
    \label{eq:3dgs}
\end{equation}
where $X$ represents a point in 3D space. 
Each voxel $V(X)$ is generated from Gaussians within a local region around it: 
\begin{equation}
    V(X|\theta_{\{i\}})=\sum_{i : ||X-\mu_i||\leq d}  G_i(X|\theta_i),
\end{equation}
Parameters of these Gaussians are optimized through successive iterations of comparing the projected image to ground truth acquisitions, interleaved with adaptive density control that creates or destroys geometry based on how well the Gaussian fits the geometry to better represent the scene. 


\noindent \uline{Loss Funcion.} We combine the projection loss $\mathcal{L}_{L2} =  \|\hat{P}_i - P_i\|_2^2$ and projected vessel centerline loss $\mathcal{L}_{clL2} = \|(\hat{P}_i - P_i)*M_{cl}\|_2^2$, and get the final loss 
\begin{equation}
   \mathcal{L}_{G} = \sum_{i}^{N} ( \alpha \mathcal{L}_{L2} + (1- \alpha ) \mathcal{L}_{clL2}),
\end{equation}
\noindent where \(\hat{P}\) and \(P\) represent the cone-beam projections of \(V\) and the actual volume, respectively. \(M_{cl}\) denotes the binary mask obtained by skeletonizing and binarizing the 2D projection images. \( \alpha = 0.5 \) set empirically. 

\subsection{Gaussian Center Predictor Training}
\label{sec:Predicting}

To provide Gaussian center initialization with coronary prior and low-noise point cloud data, we propose to train a generalizable network for estimating a rough 3D point cloud from a single view.
Specifically, we seek for a function  \( \mu = C(x) \) to predict Gaussian positions $\mu$ for a single projection $x$ using an image-to-image neural network, 
which we refer to as the GCP Network.

\noindent \uline{Gaussian Center Predictor Network.}
More precisely, this network takes the \(H \times W \times 1\) grayscale projection image as input and directly outputs a tensor of size \((H/\alpha) \times (W/\alpha) \times K\), where \(\alpha\) is the downsampling factor used to reduce the number of initial Gaussians, which can be adjusted based on the sparsity of object to reconstruct. The \(K\)-dimensional vector represents the positional parameters \(M = (d, \Delta_{x}, \Delta_{y}, \Delta_{z})\) for each Gaussian, parameterised by depth $d$ and a 3D offset $\Delta$. In practice, the network can learn to automatically predict depth of a given view, providing a rough geometry of coronary artery. The Gaussian center predictor structurally aligns with the U-Net \cite{songunet2020denoising}. The final layer is replaced by a \(1 \times 1\) convolution layer with four output channels, followed by an average pooling layer. The final output is reshaped into \(N \times 4\), where \(N\) represents the number of predicted Gaussian centers, and 4 corresponds to the positional parameters \(M\), which are transformed into the spatial positions of \(N\) Gaussians using the nonlinear activation function.

\noindent  \uline{Learning Formulation.} 
For training, we assume a multiview dataset consisting of real or simulated data. At minimum, the dataset comprises quintuplets $(P_{roj}, P_{oint}, V, d)$, where \(P_{roj}\) is a single 128 $\times$ 128 X-ray projection, \(P_{\text{oint}}\) is a point cloud composed of coronary 3D voxel positions in an \(N \times 3\) format, \(V\) represents the $128 \times 128 \times 128$ voxel data obtained from transforming the point cloud \(P_{\text{oint}}\),
and $d$ the depth map of $P_{roj}$.  
The loss function consists of three parts, each targeting a specific aspect. 

\noindent 1) \textbf{Chamfer Distance Loss.} It measures the distance between the set of predicted Gaussian center points and the set of actual label points, enabling the model to accurately localize the coronary artery. 
\begin{equation}
    \mathcal{L}_{cham} = \frac{1}{|S_1|} \sum_{p \in S_1} \min_{q \in S_2} \|p - q\|^2 + \frac{1}{|S_2|} \sum_{q \in S_2} \min_{p \in S_1} \|p - q\|^2,
\end{equation}
where \( S_1 \) and \( S_2 \) are the two sets of predicted and ground truth points. \( p \) and \( q \) are points belonging to sets \( S_1 \) and \( S_2 \), respectively.

\noindent 2) \textbf{Soft-ClDice Loss.} $ \mathcal{L}_{clDice}$ is a variant of the Dice Loss, specifically designed for evaluating tubular structure errors \cite{shit2021cldice}.
\begin{equation}
    \mathcal{L}_{clDice} = 1 - \frac{2 \times | Skel_{p} \cap Skel_{gt} |}{|Skel_{p}| + |Skel_{gt}|},
\end{equation}
where $Skel_{p}$ and $Skel_{gt}$ represent the skeletons of the predicted voxel and the ground truth voxel, respectively.


\noindent 3) \textbf{Depth Loss.} 
We employ the Scale-Invariant Logarithmic (SILog) Loss for uniformity and L1 Loss for the accuracy of predicted depth map. To obtain clear edge information, the L1 loss 
is also applied to 
the gradient space. The overall depth loss function is formulated as follows:
\begin{equation}
    \mathcal{L}_{depth} =  \mathcal{L}_{SILog} + \mathcal{L}_{GradL1} + \mathcal{L}_{MaskL1}.
\end{equation}
In more detail, the SILog loss is defined as \(\mathcal{L}_{SILog} = 10 \sqrt{var(g) + \beta (mean(g))^2}\), where \(g = \log(d_{p} + \alpha) - \log(d_{gt} + \alpha)\), and \( var \) and \( mean \) represent the variance and mean, respectively. The gradient $L_1$ loss is \(\mathcal{L}_{GradL1} = \sum \left| \nabla_x d_{p} - \nabla_x d_{gt} \right| + \left| \nabla_y d_{p} - \nabla_y d_{gt} \right|\), and the masked $L_1$ loss is \(\mathcal{L}_{MaskL2} = \sum ((d_{p} - d_{gt}) \times M_{gt})^2\). 
The final loss function for training the Gaussian center predictor is as follows:
\begin{equation}
    \mathcal{L}(C) = \gamma_1 \mathcal{L}_{cham} + \gamma_2 \mathcal{L}_{clDice} + \gamma_3 \mathcal{L}_{depth},
\end{equation}
where $\gamma_1=2 \log(\text{Iter} / 20000)$, $\gamma_2=0.5$, and $\gamma_3=0.01$ set empirically.

\begin{table}[t]
    \centering
    \caption{Quantitative comparison for 2-views coronary reconstruction with FBP, NeRP, 3DGR-CAR on ImageCAS and ASOCA. (mean ± std, best results in bold). }
    \label{tab:your_label}
    \begin{tabular}{c @{\hspace{3pt}} c|@{\hspace{5pt}} r @{\hspace{5pt}} r @{\hspace{5pt}} r @{\hspace{5pt}} r @{\hspace{3pt}}}
        \hline
        \multirow{2}{*}{Dataset} & \multirow{2}{*}{Method} & \multicolumn{2}{c@{\hspace{3pt}}}{New Projections} & \multicolumn{2}{c}{Volume} \\ \cline{3-6} 
                &        & DSC(\%) $\uparrow$ & PSNR(dB)$\uparrow$ & DSC(\%)$\uparrow$ & SSIM(\%)$\uparrow$ \\  \hline
        \multirow{3}{*}{ImageCAS} & FBP       & 33.73 $\pm$ 2.98   & 25.07 $\pm$ 2.13  & 31.36 $\pm$ 3.33    & 94.81 $\pm$ 0.76  \\ 
                 & NeRP      & 33.30 $\pm$ 3.53   & 25.34 $\pm$ 2.11  & 30.56 $\pm$ 5.87    & 95.99 $\pm$ 0.52  \\ 
                 & \uline{3DGR}    & \textbf{56.24 $\pm$ 4.21}   & \textbf{30.56 $\pm$ 2.41}  & \textbf{70.03 $\pm$ 5.95}    & \textbf{98.69 $\pm$ 0.41}   \\ 
                 \hline
        \multirow{3}{*}{ASOCA}    & FBP       & 31.58 $\pm$ 4.24    & 26.06 $\pm$ 4.35  & 40.06 $\pm$ 4.24   & 95.71 $\pm$ 1.18  \\ 
                 & NeRP      & 31.96 $\pm$ 4.25    & 26.21 $\pm$ 4.20  & 29.50 $\pm$ 7.28   & 97.66 $\pm$ 0.72  \\ 
                 & \uline{3DGR}    & \textbf{59.79 $\pm$ 5.93}   & \textbf{30.32 $\pm$ 4.43}  & \textbf{73.06 $\pm$ 6.26}   & \textbf{97.96 $\pm$ 0.61}  \\  
                 \hline 
    \end{tabular}
    \label{tab: result1} \vspace{-5pt}
\end{table}
\section{Experiments}

\subsection{Setup}
\textbf{Dataset.} Due to the unavailability and high cost of accurately calibrated, large datasets with paired X-rays and volumes, we employ digitally reconstructed radiographs (DRR) technology \cite{milickovic2000ct} to generate synthetic X-rays from multiple views. We take the 3D cone beam projection to simulate X-ray attenuation adapted from \cite{shen2022nerp} on two coronary computed tomography angiography images (CCTA) datasets. The first is ImageCAS from \cite{imagecas}. The ImageCAS dataset comprises 1000 CCTA images, of which 960 are utilized to train the Gaussian center predictor, 20 to select GCP, and 20 to test coronary reconstruction. The second is ASOCA from the MICCAI challenge \footnote{https://asoca.grand-challenge.org/}  \cite{ASOCA1,ASOCA2}. The ASOCA dataset includes 40 CCTA images, 20 used for GCP selection, and 20 for one-shot testing. Forty samples from both datasets are utilized to assess the performance of various reconstruction methods. All CCTA voxel data are resampled to \(128 \times 128 \times 128\), with the size of the projection grayscale images set at \(128 \times 128\).


\noindent \textbf{Evaluation Metrics.} The evaluation results are presented in two main aspects: first, the masked Dice similarity coefficient (DSC) and masked Peak Signal-to-Noise Ratio (PSNR) for the new reconstructed coronary projections; second, the masked DSC and Structural Similarity Index Measure (SSIM) for the reconstructed coronary volumes.  


\subsection{Comparison with Existing Methods}

\begin{figure}[t]
    \centering
    \begin{minipage}[b]{0.49\textwidth}
        \includegraphics[width=\textwidth]{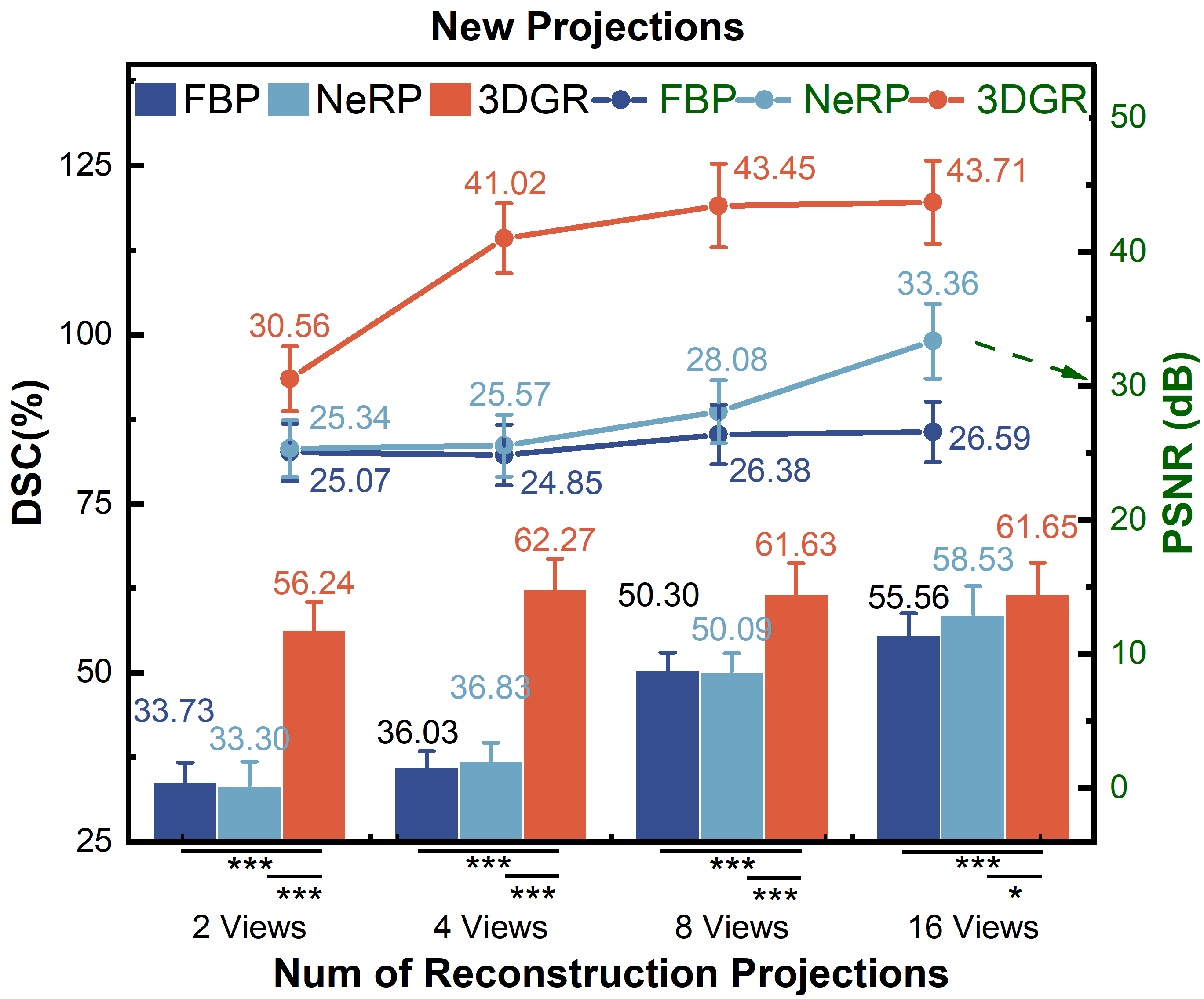}
        \caption{Comparison of the new projection quality evaluation with increasing numbers of projections on ImageCAS.}
        \label{fig:newprojs}
    \end{minipage}
    \hfill 
    \begin{minipage}[b]{0.49\textwidth}
        \includegraphics[width=\textwidth]{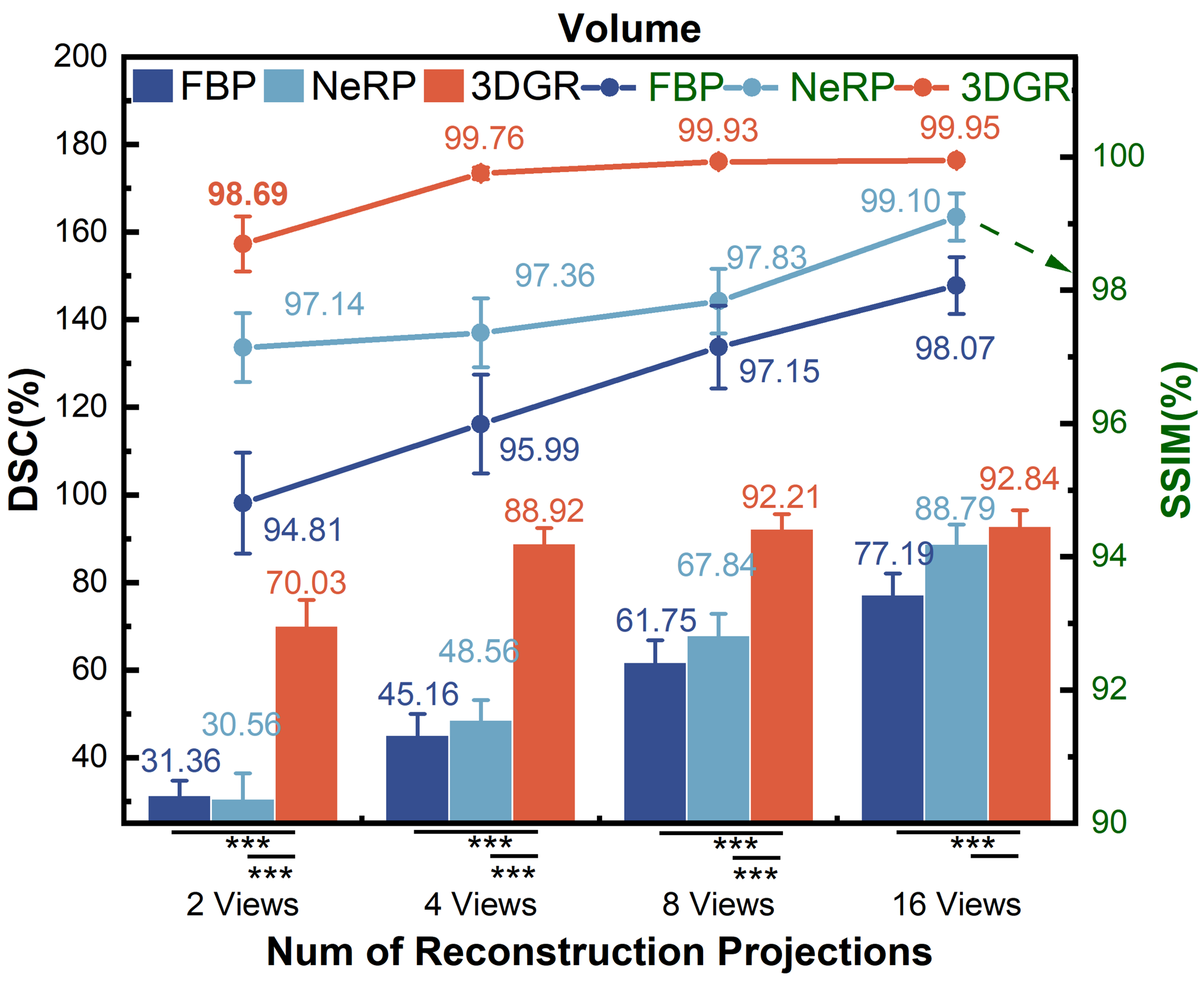}
        \caption{Comparison of volume quality evaluation with increasing numbers of projections on ImageCAS.}
        \label{fig:volume}
    \end{minipage}
\end{figure}

We conduct quantitative comparisons of FBP \cite{fbp1976reconstruction}, NeRP \cite{shen2022nerp} and 3DGR-CAR on the ImageCAS and ASOCA datasets. The 2-view coronary reconstruction results are presented in Table \ref{tab: result1}. Statistical information on the evaluation of new projections and volumes coronary reconstructions from more views (2, 4, 8 and 16 views.  The angles interval of views are $\pi$/2, $\pi$/4, $\pi$/8 and $\pi$/16.) on ImageCAS is presented in Fig. \ref{fig:newprojs} and Fig. \ref{fig:volume}, respectively. The line graph represents the PSNR and SSIM (green font), while the histogram shows the DSC (black font). Statistical significance is indicated by asterisks as follows: a single asterisk (\texttt{*}) denotes $p < 0.05$. A double asterisk (\texttt{**}) indicates $p < 0.01$, while a triple asterisk (\texttt{***}) represents $p < 0.001$.

For both ImageCAS and ASOCA, 3DGR-CAR has significantly outperformed other methods in the evaluation of newly generated projections and reconstructed voxels. It can be observed that as the number of projection views used for reconstruction increases, the discrepancy gradually diminishes from Fig. \ref{fig:newprojs} and Fig. \ref{fig:volume}. This indicates that 3DGR method possesses stronger and robuster representational capabilities in ultra-sparse coronary projection reconstruction.


\subsection{Ablation Study}

\begin{figure}[t]
    \centering
    \includegraphics[width=\textwidth]{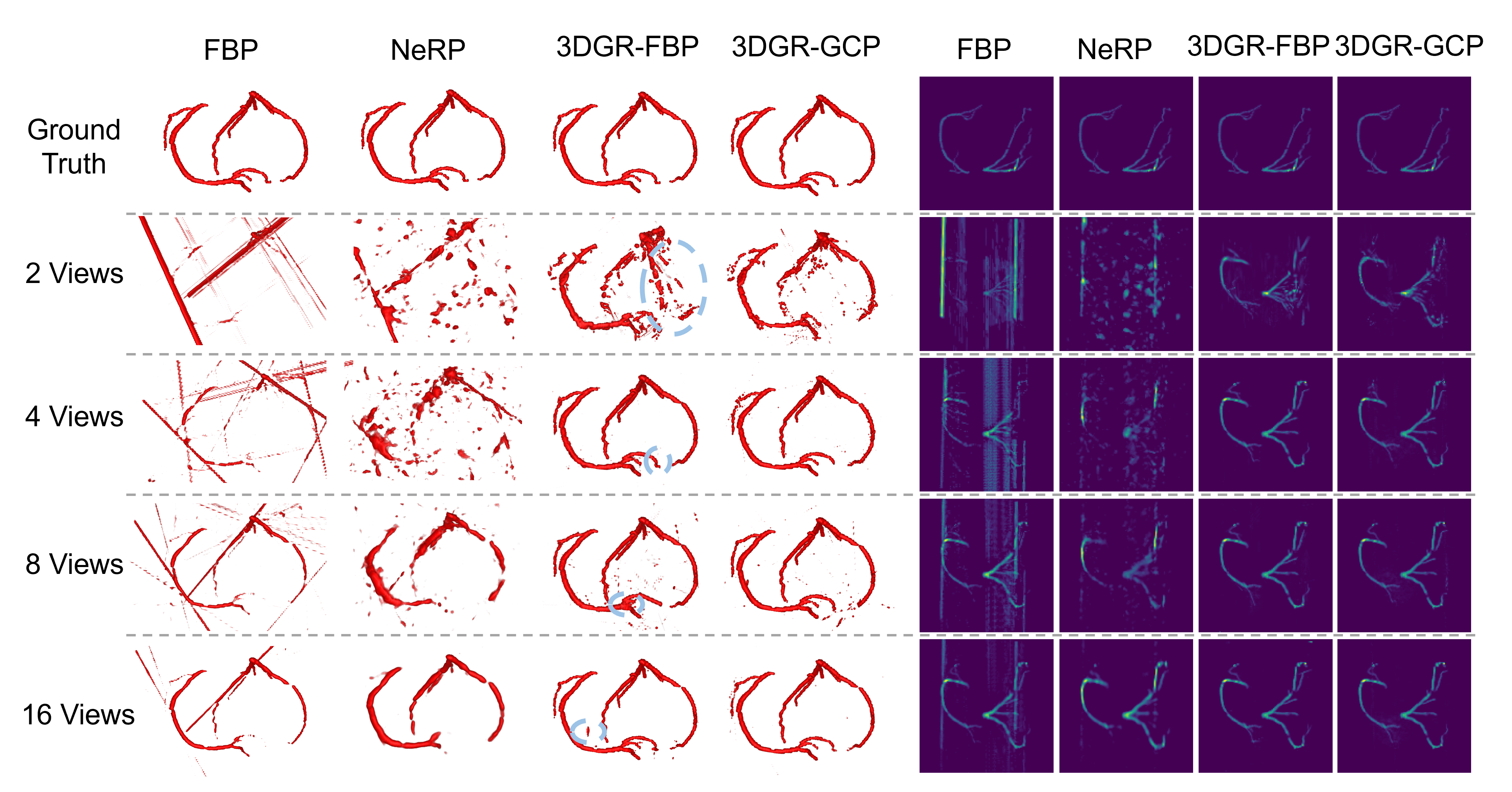}
    \caption{Comparison of 3D vascular and new projections from different methods. The blue circles marked the most differences between 3DGR-FBP and 3DGR-GCP. (left: 3D coronary reconstruction results; right: new projections of coronary arteries).} 
    \label{fig:result1}
\end{figure}

To further assess the impact of different Gaussian center initialization methods, we also compare the performance of the FBP initialization method from \cite{li2023sparse-gs} and the GCP initialization (abbreviated respectively as 3DGR-FBP and 3DGR-GCP). 
Table \ref{tab: ablation-gcp} presents the ablation study on both Gaussian initialization and the loss function Gaussian optimimization.
The 3DGR-GCP initialization method achieves superior performance on ImageCAS. This suggests that transferring prior knowledge from Gaussian center predictor can mitigate the effect of lacking 3D information to some extent, especially in extremely sparse scenarios. Also, both pixel-wise and contour-wise losses significantly impact the results.

Figure \ref{fig:result1} demonstrates that the 3DGR-based method achieves the results most similar to the ground truth in sparse coronary projection reconstruction. Furthermore, the 3DGR-GCP method outperforms the 3DGR-FBP approach under extremely sparse conditions (2-views).

\begin{table}[t]
    \centering
    \caption{Ablation Study on the Gaussian center initialization and loss function of optimization.}
    \label{tab: ablation-gcp}
    \begin{tabular}{c @{\hspace{5pt}} c|@{\hspace{5pt}} r @{\hspace{5pt}} r @{\hspace{5pt}} r @{\hspace{5pt}} r @{\hspace{3pt}}}
        \hline
        \multirow{2}{*}{Dataset} & \multirow{2}{*}{Method} & \multicolumn{2}{c@{\hspace{3pt}}}{New Projections} & \multicolumn{2}{c}{Volume} \\ \cline{3-6} 
                &        & DSC(\%) $\uparrow$ & PSNR(dB)$\uparrow$ & DSC(\%)$\uparrow$ & SSIM(\%)$\uparrow$ \\  \hline
        \multirow{2}{*}{ImageCAS}  
                 
                 & 3DGR-FBP    & 52.86 $\pm$ 4.95   & 28.90 $\pm$ 3.11  & 63.84 $\pm$ 9.73    & 97.15 $\pm$ 0.58 \\ 
                 & \uline{3DGR-GCP}    & \textbf{56.24 $\pm$ 4.21}   & \textbf{30.56 $\pm$ 2.41}  & \textbf{70.03 $\pm$ 5.95}    & \textbf{98.69 $\pm$ 0.41}   \\  \hline
                 
        \multirow{2}{*}{ImageCAS}     &$\mathcal{L}_{L2}$   & 52.36 $\pm$ 5.39     & 29.81 $\pm$ 2.35     & 63.64 $\pm$ 8.04    & 97.90 $\pm$ 00.46 \\ 
 & \uline{$\mathcal{L}_{L2} + \mathcal{L}_{clL2}$} & \textbf{56.24 $\pm$ 4.21}   & \textbf{30.56 $\pm$ 2.41}  & \textbf{70.03 $\pm$ 5.95}    & \textbf{98.69 $\pm$ 0.41}   \\ 
                 \hline

    \end{tabular}
    \vspace{-10pt}
\end{table}

\section{Conclusion}
In this paper, we introduce 3DGR-CAR, a novel 3D Gaussian Representation scheme for accurate coronary artery reconstruction from ultra-sparse 2D X-ray projections. This innovative scheme harnesses the power of 3D Gaussian representation to avoid the inefficiency in computation
, and is adeptly tailored to provide accurate initialization from ultra-sparse projections by a U-Net. Extensive experimental results on the ImageCAS and ASOCA datasets demonstrate that the proposed 3DGR-CAR significantly outperforms existing INR methods in reconstruction quality, with much shorter time. 

\begin{credits}
\subsubsection{\ackname} Supported by Natural Science Foundation of China under Grant 62271465, Suzhou Basic Research Program under Grant SYG202338, and Open Fund Project of Guangdong Academy of Medical Sciences, China (No. YKY-KF202206).

\subsubsection{\discintname}
The authors have no competing interests to declare that are
relevant to the content of this article.
\end{credits}

\bibliographystyle{splncs04.bst}
\bibliography{reference}

\end{document}